# The Measurement of the Hubble Constant $H_0$ in the Solar System

Allen Joel Anderson*

Department of Physics and Astronomy, Box 516, S-75120 Uppsala, Sweden

Abstract

This paper discusses the methodology necessary to measure the Hubble constant $H_0$ to a high degree of accuracy based upon Doppler tracking of spacecraft in the solar system. Using this methodology with available published data we determine a model independent value of the Hubble constant for the current epoch in the solar system to be $H_0$ = 2.59 ± 0.05 x $10^{-18}$ ($s^{-1}$) or as 79.8 ± 1.7 (km/s/Mpc).

We calculate the direct effect of the Cosmic Redshift on Doppler tracking of spacecraft in the solar system. It is shown that with current tracking systems, such as NASA's Deep Space Tracking Network, when the return trip light time of the Doppler signal exceeds a certain threshold, imposed by the stability of the frequency standard, the effect of the Cosmic Redshift is coherently conserved in the returning Doppler signal.

We demonstrate that in an underdetermined orbit, one determined by line of sight Doppler alone, that if this Cosmic Redshift term is not accounted for, the orbit determination program (ODP) miscalculates the actual recessional velocity of the spacecraft from the measured recessional velocity causing a mismatch between the actual and the predicted trajectory of the spacecraft. One consequence is that the ODP will generate Doppler residuals, the difference between the actual trajectory and the predicted trajectory which show an anomalous force. When this effect is integrated in long arc solutions, it can grow to considerable magnitude. We show that the ODP uniquely separates the Cosmic Redshift term from velocity Doppler sources and that the solution can provide an accurate determination of $H_0$.

Introduction

A major effort of astronomical research is to estimate the value of the Hubble constant $H_0$ over a wide range of distances. Figure 1 is a summary graph of the estimate of $H_0$ from the 2001 HST Key study (1). While advances in new technologies and methodologies allow the measurement of $H_0$ to greater and greater distances, the ability to measure $H_0$ accurately at very short distances has not been investigated. This may be due to the fact that traditional astronomical methods allow for an estimation of $H_0$ only at distances greater than several Mpc.

*email: allen.joel.anderson@gmail.com, allen.j.anderson@fysast.uu.se

In a recent paper, Riess, et.al. (2) stress the need to determine the zero point value of $H_o$ for the Cepheids in order to define better the value of $H_o$ at all distances. Figure 2 summarizes the measurement of $H_o$ for the Cepheids from the HST study (1). At the moment there is no short range model independent value of $H_o$.

As we will discuss, celestial mechanics experiments are currently sufficiently accurate that the effect of the Cosmic Redshift causing a spectral shift in the Doppler tracking of spacecraft is measurable. If the direct effect of a Cosmic Redshift on the Doppler measurement of spacecraft is not accounted for, the actual measurement of the Cosmic Redshift is either placed in other available parameters being fit by the orbit determination program (ODP) or left as a badly fit trajectory solution leading to the appearance of misfit Doppler residuals and anomalous forces.

Literature currently reports instances of anomalous and badly fit trajectories of spacecraft tracked by NASA's Deep Space Tracking Network (DSN) (3), (4), (5), (6) but the analysis of the direct effect of the Cosmic Redshift on the Doppler signal itself and how this would impact the predicted trajectory produced by the ODP has not been well described. These reported spacecraft anomalies generate discussion among a wide portion of the scientific community, but leave us with a less than adequate understanding of a solution to the problem (7).

The Cosmic Redshift, the Doppler Tracking Signal and the Hubble Constant

Celestial mechanics has traditionally relied upon measurements of positions of planets measured against the background of stars. Together with GM based equations of motions, they provide the basis for the celestial reference frame. In earlier measurements, no electromagnetic data of spectral measurements was utilized. More recently, data from Doppler tracking of spacecraft, basically a spectral measurement of electromagnetic radiation, places spacecraft into a solution of the celestial reference frame. While traditional celestial mechanics has no need to incorporate the Cosmic Redshift (the FLRW metric causing a spectral shift in electromagnetic radiation), Doppler tracking does. Although it is stated that current orbit determination programs are fully relativistic, they do not, in fact, incorporate this implied FLRW redshift correction to spacecraft Doppler measurements.

The FLRW metric for the Cosmic Redshift in its simplest form can be written as follows:

Equ. 1     $1+ z = a(now)/ a(then)$

Here a is the FLRW metric scale factor. The ratio of this term in seconds of time and its effect on z provides a measure of the magnitude of $H_o$. z is an electromagnetic spectral shift in wavelength and is directly dependent upon the difference in seconds in time between two measurements: that is, the difference between two space metrics at different times from each other. This difference is a purely spectral shift, representing the photon accommodating to a change in the space metric. Thus, $H_o$ is simply a scale factor of the rate at which the photon changes its wavelength in seconds of time.

From this the Cosmic Redshift effect for a fractional change of frequency of an electromagnetic signal in time is:

Equ. 2     $-\Delta f/f = t \times H_o$

Here t is the time difference between send and receive in seconds of time, which is equivalent to the value of the return trip light time (RTLT) of the DSN. $H_o$ used here is the standard approximate value of: $H_o = 2.5 \times 10^{-18}$ (s$^{-1}$).

A successful measurement of this fractional change of frequency would yield a precise value of $H_o$ at the current epoch. In fact, this would be the intent of any experiment to measure $H_o$ in this manner.

Figure 3 plots this relationship together with an estimate of the DSN's ability to measure fractional frequency Doppler shifts. Since the 1970s (8) the DSN's ability to make this measurement is confined by the stability of the frequency standard, measured as Allen variance in time, and is displayed here (9). In addition, this basic limit shows that for a RTLT greater than a certain fraction of an AU, currently around 0.1 AU, the Cosmic Redshift is incorporated into the Doppler signal and conserved in the tracking data. In other words it impacts the measured Doppler signal.

For most spacecraft, disturbances of several types reduce the ability to model the trajectory sufficiently well. If one does not have the opportunity to monitor an unperturbed trajectory over long periods of time, where the spacecraft is removed from such effects as solar radiation pressure and other variable parameters for which the ODP might easily absorb the Cosmic Redshift effect, the Cosmic Redshift has an effect on the trajectory solution but it is difficult to quantify. However if these limitations can be overcome, such as with a single axis spin stabilized spacecraft in the outer solar system, the effect of the Cosmic Redshift should become apparent. Then: How would the ODP, not programmed to model this Cosmic Redshift term, produce a trajectory solution and what would this solution look like?

The ODP trajectory solution

Before attempting a solution we propose a "Gedanken" experiment in order to better understand the nature of the trajectory problem. In this "Gedanken" experiment we will assume that the Cosmic Redshift takes the form as described above, and we will assume that the ODP attempts to model the spacecraft trajectory given no apriori information of its existence and wherein the ODP has no available parameter in which to fit the Cosmic Redshift component of the Doppler data.

To begin with we must consider how the DSN generates a raw Doppler tracking measurement. This raw Doppler value is the basis for all future results and in a chain of reductions that eventually produce a Doppler residual (misfit results produced by the ODP).

Figure 4 provides a simplified schematic of the major components of the Doppler tracking measurement for the DSN tracking system (10). In the figure one sees the major components together with several potential sources of disturbance in the tracking data. The DSN produces a coherent closed loop Doppler measurement, meaning that the DSN coherently keeps the phase of the frequency shift for each second of measurement throughout its entire tracking pass. In reality this produces a Doppler measurement of (Hz/s per s), or (cycles per s/s²), which has the dimension ($s^{-1}$).

Moreover, this Doppler tracking measurement coincides with the same dimensional measurement as the Hubble constant $H_o$ in ($s^{-1}$). In this sense one may describe the DSN and the manner in which it produces its closed loop Doppler data as a perfectly designed Hubble constant measurement instrument.

The raw Doppler measurement from the DSN is the sum of all Doppler components; therefore, all the sources of Doppler shifts from the spacecraft are included in this data, and they must be distinguished and carefully eliminated through various procedures. To do this the ODP attempts to place the spacecraft in a rest frame solution in the celestial reference frame thus producing a Doppler residual of value zero. However, in this case the ODP does not have sufficient information to correctly solve this problem. In addition other types of corrections effecting the electromagnetic signal, such as the effects of the solar plasma and its variation in time are needed and must be modeled before the Hubble constant can be accurately estimated. This estimation is the challenge of our "Gedanken" experiment, and it requires great care to correct for only those Doppler sources necessary while not adding imaginary sources, so as not to corrupt the final result. Finally the "Gedanken" experimenter must try to understand what the ODP does with the Cosmic Redshift term for which it has no parameter.

It is important to note that the ODP uses all the Doppler measurement data to produce its estimation of the orbital elements of the spacecraft. If the ODP solution is based upon measurements from the line of sight Doppler data alone, the solution will be underdetermined, as components at right angles to the line of sight (off axis components) are not well determined. One can say, then, that the ODP solution itself is but a 'pseudo' solution to the orbital elements of the spacecraft. The underdetermined ODP orbit solution will include the non gravitational (spacecraft self generated) force term and the Cosmic Redshift term, as the ODP has no means of separating out these terms.

Thus for our "Gedanken" analysis: As long as all the sources of Doppler data that the ODP uses is a real velocity Doppler, the underdetermined ODP orbit solution will fit the actual trajectory regardless of the Doppler source. However, if one of the sources is not a real velocity Doppler, in this case the Cosmic Redshift term, then the underdetermined ODP orbit solution will not fit the actual trajectory. This concept is vital to understanding the true nature of the Cosmic Redshift term in the underdetermined ODP orbit solution. The Cosmic Redshift term stands out from all other sources of Doppler so long as all the other sources produce an actual velocity on the spacecraft. The Cosmic Redshift term is unique from all the rest of the Doppler sources as it does not effect the actual trajectory of the spacecraft. It only effects a frequency shift of the returning photon. Thus the Doppler residual which is produced is the difference of the actual trajectory minus the underdetermined ODP orbit solution plus whatever stochastic terms remain in the Doppler data.

To distinguish whether or not the Cosmic Redshift is a part of the ODP orbit solution, we make the following postulate: (1) From equation 2 we will assume that the Cosmic Redshift term is included the DSN Doppler data. This is consistent with the FLRW metric. (2) Therefore either the ODP has included the Cosmic Redshift term in its orbit solution or it has not. (3) If the ODP orbit has not included the Cosmic Redshift term in its solution, then the residuals, that is the difference between the actual trajectory and the ODP orbit, will show a positive acceleration away from the observer for the Cosmic Redshift term. This is because the Cosmic Redshift term is neither in the actual trajectory nor the ODP solution, and will be seen only in the residual which is in the DSN Doppler data. (4) On the other hand, if the Cosmic Redshift term is included in the ODP solution, then the residual will be equal in magnitude to the Cosmic Redshift term with reverse sign. This is because the Cosmic Redshift term is not in the actual trajectory, but is included in the ODP solution, while by inference the spacecraft non gravitational (spacecraft self generated) force term will be included in both the actual trajectory and the ODP trajectory and thereby be eliminated in the residuals, as this term cannot be separated out from an underdetermined ODP orbital solution.

We can describe the two alternatives of the postulate in the following way and we define:

equ. 3    $D(total) = D(cm) + D(ng) + D(H_o) + D(sto)$

where $D(total)$ = total Doppler composed of: $D(cm)$ the JPL/IAU celestial mechanics component; $D(ng)$ the non gravitational (spacecraft generated) component; $D(H_o)$ the $H_o$ component; and $D(sto)$ the stochastic component. Here we define the non gravitational component as any magnitude and rate change caused by the spacecraft generated forces. The stochastic component is composed of plasma noise, system noise and other short term or sinusoidal features not included in the other 3 terms.

We will distinguish between a well determined, sometimes referred to as an overdetermined ODP orbital solution, and an underdetermined ODP orbital solution, one dealing with line of sight Doppler only.

1) When the orbit is a well determined ODP solution ,$\psi$, can be written as:

equ. 4    $\psi = D(cm)$

Then the Doppler residual, $D(res)$, is defined as the actual trajectory, $\zeta$, minus the well determined ODP solution, $\psi$, plus any remaining terms:

If the actual trajectory, $\zeta$, matches the well determined ODP solution exactly then $D(ng) = 0$ and $\zeta = \psi$, and the resulting Doppler residual is:

equ. 5    $D(res) = \zeta - \psi + D(H_o) + D(sto) = + D(H_o) + D(sto)$

Now if D(ng) ≠ 0, then the actual trajectory, ζ, does not match the well determined ODP solution exactly and ζ = D(cm) + D(ng), and the resulting Doppler residual is:

equ. 6    D(res) = ζ – ψ + D($H_o$) + D(sto) = + D(ng) + D($H_o$) + D(sto)

This is the Doppler residual we would expect to find for a well determined ODP trajectory solution.

2) However when the orbit is underdetermined, one determined by line of sight Doppler alone, the ODP includes additional Doppler terms D(ng) and D($H_o$) in its underdetermined orbit solution, as the ODP is unable to separate out these terms in the line of sight Doppler data. We then have:

equ. 7    ψ = D(cm) + D(ng) + D($H_o$)

And the actual trajectory, ζ, can be written:

equ. 8    ζ = D(cm) + D(ng)

Then the Doppler residual, D(res), is defined by the actual trajectory, ζ, minus the ODP underdetermined orbit solution, ψ, plus any remaining terms:

equ. 9    D(res) = ζ – ψ + D(sto) = - D($H_o$) + D(sto)

This, then, is the Doppler residual we would expect to find in an underdetermined ODP orbital solution.

Figure 5 plots the above giving 3 examples. These examples show what we believe happens with the Cosmic Redshift term as the ODP estimates the spacecraft orbit. The X-axis is time and the Y-axis is a projection along the line of sight Doppler. Example 1 sets the magnitude of the spacecraft non gravitational (self generated) force equal to the equivalent Cosmic Redshift term, $H_o$, with the same sign. Example 2 sets the magnitude of the spacecraft non gravitational (self generated) force equal to the equivalent Cosmic Redshift term, $H_o$, but with opposite sign. Example 3 sets the magnitude of the spacecraft non gravitational (self generated) force equal to zero. From equation 9 and figure 5 the "Gedanken" experiment demonstrates that regardless of the magnitude or sign of the spacecraft non gravitational (self generated) force the ODP orbital solution includes the Cosmic Redshift term (for which it has no parameter) producing a Doppler residual of equivalent value with reversed sign.

Although at first this result may seem counter intuitive, it is important to remember that the ODP sees the Cosmic Redshift term, $H_o$, as the equivalent of a constant Doppler velocity increment. The ODP calculates its orbit including this misinterpreted Cosmic Redshift term as well as the Doppler produced by the spacecraft non gravitational (self generated) force and any rate change it has. The ODP has no information to do otherwise in an underdetermined orbit calculation.

Thus we have an interesting question: What is the form that the Cosmic Redshift term now takes? Since the Doppler data produced by the DSN has the form of the Hubble constant with dimension ($sec^{-1}$), the Cosmic Redshift term, $H_o$, will not be dependent on the distance to the spacecraft, but will be seen in an underdetermined ODP orbit residual as a constant positive frequency shift (that is reverse sign) as long as $H_o$ remains constant over the distances being measured. We now have a good indication of what to look for when searching for the Cosmic Redshift term with real spacecraft Doppler residuals.

Since the ODP uses all the Doppler data, the "Gedanken" experiment now has only to consider the effects of the path length corrections on the electromagnetic signal as it passes from the earth to the spacecraft and back. These are the plasma and troposphere corrections shown in figure 4. The plasma corrections, which can be large for long path lengths, are frequency dependent, and therefore on a two frequency link, such as an S and X band link, or a X and Ka band link, the plasma effects are measurable and correctable. If only a single frequency link is available, then a model reduction is needed. For a single link correction we can use a two link solution obtained from other spacecraft to provide a model for the single link correction. The troposphere corrections have very much less impact than the plasma corrections. They comprise a short term and long term seasonal correction (11). Potentially, they can improve the analysis of long arc solutions to the trajectory, but their influence is small.

This completes the "Gedanken" experiment.

The search for the Cosmic Redshift

We believe the "Gedanken" experiment reveals sufficient information to solve the problem of the measurement of the Cosmic Redshift in the solar system. First we review earlier papers written regarding tracking anomalies about spacecraft to see if they provide sufficient information in their analysis so that one can deduce the Cosmic Redshift term, $H_o$.

Paper (3) is not sufficient. We see that paper (4) has sufficient detail to potentially deduce the Cosmic Redshift term and it is a thoroughly written work of a complex scientific experiment. We observe that the reported Doppler tracking anomaly has a positive frequency shift, that is in the opposite direction to the Cosmic Redshift term. From our postulate and from equations 7 thru 9 this is strong indication that the ODP has incorporated the Cosmic Redshift term in its underdetermined orbit solution and that the Doppler residual represents $H_o$ with reverse sign. We observe what went wrong in the authors analysis of the Doppler tracking anomalies.

A conceptual mistake of the paper is that it proceeds as if the ODP Doppler anomaly is a real force on the spacecraft. It rejects evidence that the anomaly is produced by a frequency shift in the data. While it makes several attempts to tie the anomaly to frequency shifts, it always rejects them. First it proposes a common drift in the frequency standards to account for the Doppler anomaly. But it rejects this on the grounds that it is unlikely that all the frequency standards used by the DSN would drift in the same manner. Second the paper proposes a quadratic in time adjustment to the data and calls it an "expansion of space" model. This proposal is rejected on grounds that the ODP "doesn't prefer" this solution when not considered as an 'apriori' solution. Here I quote from paper (4) page 47:

> " The orbit determination process clearly prefers the constant acceleration model, ap, over the quadratic in time model a (quad). This implies that a real acceleration is being observed and not a pseudo acceleration. We have not rejected this model as it may be too simple in that the motions of the spacecraft and the Earth may need to be included to produce a true expanding space model. Even so, the numerical relationship between the Hubble constant and ap, which many people have observed, remains an interesting conjecture."

The paper proceeds with the assumption that one is dealing with a real force on the spacecraft and attempts to create a "drag free" trajectory based upon a correction to the non gravitational (spacecraft self generated) forces and it applies this correction to the data. Meanwhile, it observes that these estimated spacecraft forces, which change over the duration of the 11 ½ year long arc solution studied, in fact produce no observable effects on the anomaly. The paper does not understand what this implies. Here I quote from paper (4) page 34 that makes a key observation:

> "Finally, we want to comment on the significance of radioactive decay for this mechanism. Even acknowledging the Interval jumps due to gas leaks, we reported a one-day batch sequential value for ap = $(7.77 \pm 0.16) \times 10^{-8}$ cm/s². From radioactive decay, the value of ap should have decreased by 0.75 of these units over 11.5 years. This is 5 times the above variance, which is very large with batch sequential. Even more stringently, this bound is good for all radioactive heat sources. So, what if one were to argue that emissivity changes occurring before 1987 were the cause of the Pioneer effect? There still should have been a decrease in ap with time since then, which has not been observed."

This is a key observation. It confirms our postulate. While the ODP uses these real spacecraft generated forces in its one-day batch sequential solution in its predicted trajectory, they do not impact the Cosmic Redshift term. It is the Cosmic Redshift term that produces the 'anomaly', nothing more. The Cosmic Redshift term is intact and singled out in the underdetermined ODP one-day batch sequential orbit solution as the only Doppler component that does not produce a real velocity on the spacecraft, thereby producing a Doppler residual just as our "Gedanken" experiment in equations 7 thru 9 has described.

Having singled out the Cosmic Redshift term, we potentially can measure $H_o$. First we check to see that this 11.5 year one-day batch sequential solution corrects for the solar plasma for the single frequency S band Doppler data of this spacecraft. We find on page 24 that it does. (See table I from paper (4) and the caption within). The data used to provide this model for the correction of the S band data is from the recent Cassini mission which uses X and Ka band data that provides an accurate solar plasma model. We note that there will be some additional error included in this correction because it does not cover the same time period, but it is a sufficiently accurate model. We are now prepared to make an estimate of $H_o$ from Doppler tracking of spacecraft in the solar system.

Discussion of an estimate of the Hubble Constant and the error terms

Figure 6 taken from paper (4) shows the trajectories of the Pioneer and Voyager spacecraft in the solar system projected onto the ecliptic plane. The tracking data used for this study is for the period of Pioneer 10 from 1987 to 1999 which cover a distance range of approximately 40 to 90 AU.

The Cosmic Redshift term emerges in the data as the solar radiation pressure (SRP) on the spacecraft decreases to such a level that the redshift term becomes apparent. It is clear that the Cosmic Redshift term has been present earlier in the trajectory, but has been masked by the SRP. Figure 7 taken from paper (4) illustrates this in a drawing showing the Doppler residuals from both Pioneer 10 and 11 plotted as circles together with an estimate of the SRP and showing the combined effect of the two terms as the Cosmic Redshift term emerges from the data.

The paper attempts to understand the nature of this anomaly, finally attributing it to an "anomalous" force. It also notes that tracking data from other spacecraft, specifically Galileo and Ulysses, appear to have a similar effect, though more difficult to quantify. It even attempts to fit its solutions of the "anomaly" to these spacecraft "anomalies" with some success. It provides numerous solutions for different periods of the data using two independent ODPs and gets similar results. Much of the analysis, however, involves discussion of the spacecraft non gravitational (self generated) forces. This is unfortunate as this has little to do with a solution to the problem, as we have shown from our analysis in equations 1 through 9.

The paper observes both an annual and diurnal term which appears in the data. Figure 8 taken from paper (4) illustrates these terms from a detail portion of this data from 1996. The implication here is that the orbit determined for Pioneer 10 is in error by a large amount. This is because these two sinusoidal terms, associated with the position of the Earth in its orbit and the positions of the tracking stations on the surface of the Earth are much better known in the celestial reference frame than these errors would suggest. The error lies not with the Earth, but with the determined spacecraft position produced by the ODP solution. The easiest way to interpreted this is an out of the ecliptic orbit error of the spacecraft projected onto the tracking stations on the Earth. This is a sign of a large underdetermined orbit error of the spacecraft in the off axis line of sight Doppler tracking direction.

The paper comments on this. It discusses that navigation was performed using only line of sight Doppler leading to an underdetermined orbit solution which can account for the annual and diurnal terms in the tracking data. However the paper fails to understand the full impact of this observation on the understanding to the solution of the measured "anomaly". Here I quote from page 42 of the paper:

" We conclude that both for Pioneer 10 and 11 there are small periodic errors in the solar system modeling that are largely masked by maneuvers and by the overall plasma noise. But because these sinusoids are essentially uncorrelated with the constant ap, they do not present important sources of systematic error. The characteristic signature of ap is a linear drift in the Doppler, not annual/diurnal signatures."

We now turn to the error estimates that paper (4) has studied. These are summarized in table II of paper (4). Table II consists of 3 categories: (1) Systematics generated external to the spacecraft, (2) Onboard generated systematics, and (3) Computational systematics. The values given are in units $10^{-10}$ m/s².

The largest of these error categories is (2) with an estimated bias of nearly 0.90 with a quadrature sum uncertainty of nearly ± 1.30. The error estimation in this category has been used by each of the many papers published for over a decade to modify the actual measured Doppler residual of the anomaly. As we have shown in equations 7 through 9 this is incorrect, as the category (2) Onboard generated (non gravitational) systematics have been eliminated from the Doppler residuals by the ODP in its underdetermined orbit solution and no correction using this term should be applied to the anomaly to calculate the Cosmic Redshift term. This has caused the largest obstacle in the interpretation of the true nature of the anomaly. As observed the reversed sign for $H_o$ is characteristic of an underdetermined orbit solution.

The second largest of these error categories is (3) with no bias term but with a quadrature sum uncertainty of nearly ± 0.40. This term is primarily based upon the annual/diurnal sinusoids (shown in figure 8). These have been largely dealt with in the 1 day batch sequential solution estimation of the uncertainty and therefore will not be considered as additional uncertainty here. As commented upon by the authors of paper (4), they do not effect the overall value of the constant frequency shift term. A least squares analysis of the data itself could potentially remove this error from any future solution.

The smallest error category is (1) with a potential bias of 0.03 with an uncertainty of ± 0.06. This error essentially is based upon solar radiation pressure, solar corona and Kuiper belt's gravity, the bias coming from the solar radiation pressure term. These are not dealt with by the analysis of the data itself, being model dependent, however they are very small with a potential bias of 3.9 parts per thousand and an uncertainty of 7.7 parts per thousand on the measured value of the anomaly of 7.77 x $10^{-10}$ m/s².

The best overall solution for the anomaly, which paper (4) discusses in some detail, is the 1-day batch sequential analysis for the entire 11 ½ year data set and which has been referred to earlier in the quote from page 34 of the paper. Figure 9 taken from paper (4) illustrates an example of the stochastic variance in the data from this solution. This is referenced to the adopted IAU Ephemeris (JPL DE403/LE403).

We use in our analysis the value of $H_o$ obtained from the 11 ½ year 1 day batch sequential solution. As discussed the underdetermined ODP solution separates out the non gravitational (spacecraft generated) force while the stochastic term in the data provides a realistic error estimate associated with the Cosmic Redshift term. This value is (7.77 ± 0.16) x $10^{-10}$ m/s². We consider that this 1 sigma error estimate encompasses all values obtained between 7.61 to 7.93 x $10^{-10}$ m/s² which appear in column 3, table I, of paper (4) containing solutions from two independent ODP programs, JPL/sigma and CHASMP. This value equates to a measured value of the Hubble constant in the solar system of $H_o$ = 2.59 ± 0.05 x $10^{-18}$ ($s^{-1}$) or as 79.8 ± 1.7 (km/s/Mpc).

Thus we conclude with an estimate of the Hubble Constant for the current epoch in the solar system.

Implications of these results

This paper proposes a solution to the spacecraft tracking anomalies which have been reported. This solution is based upon well established properties of the FLRW spacetime metric and the manner in which the ODP produces its Doppler residuals.

The estimated value for $H_o$ in the solar system from this study, assuming $H_o$ is very nearly constant over these distance ranges, implies that we may need to raise the current zero point estimate of $H_o$ for the Cepheids to between 78.1 and 81.5, with a likely value around 79.8 (km/s/Mpc). This is about a 6% increase from the current estimates of 75 (km/s/Mpc), but well within the 1 sigma error estimates as given in figure 2 for the Cepheids.

The value of $H_o$ obtained here could be viewed as being more in accord with the measured results obtained from the entire HST Key study for estimates of $H_o$ at distances less than ≈ 60 Mpc. It should be pointed out that in this distance range the spread of the values in the HST study is extremely large, however the mean of the ten values at less than ≈ 42 Mpc from the HST study, figure 1, is 79.3 (km/s/Mpc). This is just 0.7% from its estimated value here of 79.8, and within the 1 sigma error bars of both studies.

Figure 10 shows a detail from the HST Key Study for estimates for Ho for distances less than ≈ 60 Mpc together with the spacecraft solar system determined value plotted in yellow. The reported 'tension' between these close range estimates of $H_o$, as discussed by Riess, et.at. 2009 (2), lies between these values and those at much greater range.

A paper by Tegmark, et.al., 2006, (12) describes a 12 parameter λCDM model solution using data from the SDSS of luminous red galaxies (LRGs) constrained with data from WMAP3 and the HST Key Study. Figure 16 from that paper is reproduced here as figure 11 showing the allowed region for the Hubble parameter, h, defined as $H_o$/(100 km/s/Mpc), plotted against Ωtot, the ratio of total mass energy density to critical density for their model solution. Superimposed in green on their figure 16 in the allowed region is the estimate of $H_o$ from Spacecraft Doppler tracking using a 2 sigma error estimate consistent with the 95% confidence of the figure. In their model at Ωtot=1, the green region for this estimate of $H_o$ is between 76.4 and 77.5 at the 2 sigma level.

For a summary of many current issues regarding $H_o$ and its measured uncertainties, see for example, Jackson, 2007 (13).

A separate implication of this study applies to the Doppler tracking of spacecraft itself. Assuming this solution is correct, it would seem imperative to include the effect of the FLRW metric in future orbit determination programs. Without this modification one should expect to see continued tracking anomalies.  A comparison might be made between the use of Doppler tracking and the use of the magnetic compass for navigation. In earlier centuries, without corrections of the magnetic deviation from true north over the routes traveled by seagoing mariners, serious navigational errors occurred. Similarly, without corrections of the FLRW metric to Doppler tracking, serious navigational errors occur.

Conclusions

This work describes a solution for the measurement of $H_o$ in the solar system in accordance with the FLRW spacetime metric. The measurement obtained in this study for the Hubble constant is dependent upon the particular experimental limitations of the tracking data. These results are in agreement with both theoretical and experimental results obtained over a wide range of theory and experiment in physics. If this interpretation is incorrect, this would imply that both theoretical and experimental results from a wide range of physics would have to be modified in some extraordinary way.

The Cosmic Redshift has never been measured at this very close range. The value obtained here for the solar system is $H_o = 2.59 \pm 0.05 \times 10^{-18}$ (s$^{-1}$) or as 79.8 ± 1.7 (km/s/Mpc) . The next closest estimates of $H_o$ are at distance ranges exceeding $10^{10}$ times greater, each individually having measured uncertainties roughly 10 times larger than this measurement.

Considering this solution to be correct indicates that Doppler tracking of spacecraft provides a method for a highly accurate determination of a model independent value of $H_o$ at very short range. The value of $H_o$ obtained in this study has been accomplished using a single frequency S band Apollo era spacecraft. Using modern spacecraft specifically designed to measure $H_o$, an order of magnitude improvement on the accuracy of this result could be expected with the potential to measure the value of $H_o$ to within 0.2% uncertainty. These results would strongly suggest that future orbit determination programs need contain algorithms to identify and to estimate the Cosmic Redshift term, and that spacecraft missions, current and planned, should consider Doppler tracking methods to determine $H_o$ as an important experimental measurement goal.

Acknowledgments

I would like to thank members of the Department of Physics and Astronomy at Uppsala University, particularly Prof. Nils Bergvall for his continued interest and encouragement that led to the writing of this paper, to Prof. Bengt Gustafsson for his inspiration and friendship for these many years, and to Jeff Jennings for providing important improvements to this manuscript. I extend a very special thanks to John Anderson, mentor, colleague, friend, for his devotion to Pioneer 10 and his dedicated work over many years and to the entire Pioneer 10 team from a bygone era who made possible this exceptional measurement.

References

1. Freedman, W., et. al., Final Results from the Hubble Space Telescope Key Project to Measure the Hubble Constant, ApJ 553,47.[arXiv:astro-ph/0012376v1 (18 Dec 2000)].

2. Riess, A., et. al., A Redetermination of the Hubble Constant with the Hubble Space Telescope from a Differential Distance Ladder, ApJ 699.539. [arXiv:0905.0695 (1 May 2009)].

3. Anderson, J. D., et. al., Indication, from Pioneer 10/11, Galileo, and Ulysses Data of an Apparent Anomalous, Weak, Long-Range Acceleration, Phys. Rev. Lett., 81, 2858-2861. [arXiv:gr-qc/9808081v2 (1 Oct 1998)].

4. Anderson, J. D., et. al., Study of the anomalous acceleration of Pioneer 10 and 11, Phys. Rev. D., 65(8), 082004. [arXiv:gr-qc /0104064v5 (10 Mar 2005)].

5. Nieto, M.M. and J. D. Anderson, Using Early Data to Illuminate the Pioneer Anomaly, Clas. Quant. Grav., 22, 5343-5354. [arXiv:gr-qc /0507052v2 (4 Oct 2005)].

6. Turyshev, S. G., and V. T. Toth, The Puzzle of the Flyby Anomaly, [arXiv:0907.4184v1 [gr-qc] (23 Jul 2009)].

7. Turyshev, S. G., and V. T. Toth, The Pioneer Anomaly, [arXiv:1001.3686v1 [gr-qc] (20 Jan 2010)].

8. Anderson, A. J., Detection of Gravitational Waves By Spacecraft Doppler Data, in Experimental Gravitation, B. Bertotti (ed.), p.235-246, Accademia Nazionale Dei Lincei, Rome (1977).

9. Armstrong, J. W., Low-Frequency Gravitational Wave Searches Using Spacecraft Doppler Tracking, Living Rev. Relativity 9, (2006), 1 (revised 15 January 2008).

10. Hellings, R.W., et. al., Spacecraft Doppler Gravity Wave Detection, Phys. Rev. D., 23: 844-851 (1981).

11. Anderson, A.J., Measured Path Length Variations Due to Changes in Tropospheric Refraction, in Refractional Influences in Astrometry and Geodesy, E.Tengström & G.Teleki (eds.), p.157-162, IAU Symposium #89, D.Reidel (1979).

12. Tegmark, M., et.al., Cosmological Constraints from the SDSS Luminous Red Galaxies, Phys. Rev. D., 74: 123507. [arXiv:astro-ph/0608632v2 (30 Oct 2006)].

13. Jackson, N., The Hubble Constant, Living Rev. Relativity 10, (2007), 4 [arXiv:0709.3924v1 [astro-ph] (25 Sep 2007)].

FIGURES

Figure 1.   $H_o$ from the Hubble Space Telescope Key Study, all data (from ref. (1)).

Figure 2.   $H_o$ for the Cepheids (from ref. (1)).

Figure 3.   The effect of the FLRW metric on DSN Spacecraft Doppler tracking data.

Figure 4.   Simple schematic of the DSN spacecraft Doppler tracking system (from ref. (10)).

Figure 5.   Three Examples showing how the Doppler residuals for the measurement of $H_o$ are independent of the spacecraft generated force in a underdetermined ODP orbit solution.

Figure 6.   Schematic showing the trajectories of the Pioneer 10 & 11 and Voyager 1 & 2. Tracking Data used in the study is from the years 1987 to 1999 extending into a period not shown in this figure. (figure 6 of paper (4)).

Figure 7.   Diagram showing the "anomalous" force emerging in the Doppler residuals as the solar radiation pressure (SRP) term is diminished. (figure 7 of paper (4)).

Figure 8.   Detail of the Doppler residuals from 1996 showing the annual and diurnal terms in the data. (figure 18 of paper (4)).

Figure 9.   Example of the acceleration residuals of the 11 ½ year Pioneer 10 data set from the ODP 1-day batch sequential analysis. (figure 17 of paper (4)).

Figure 10.  Detail showing the estimates of $H_o$ to ≈ 60 Mpc from the entire HST Key study plotted with the spacecraft result from the solar system. (from ref. 1 and from this paper).

Figure 11. The allowed region for the Hubble parameter, h, plotted against $\Omega_{tot}$ for the λCDM model solution from the Tegmark, M, et.al. 2006 study with 95% confidence constraints. The spacecraft result from the solar system using a 2 sigma error estimate consistent with 95% constraints of the figure is marked in green. (from figure 16 of ref. (12) and from this paper).

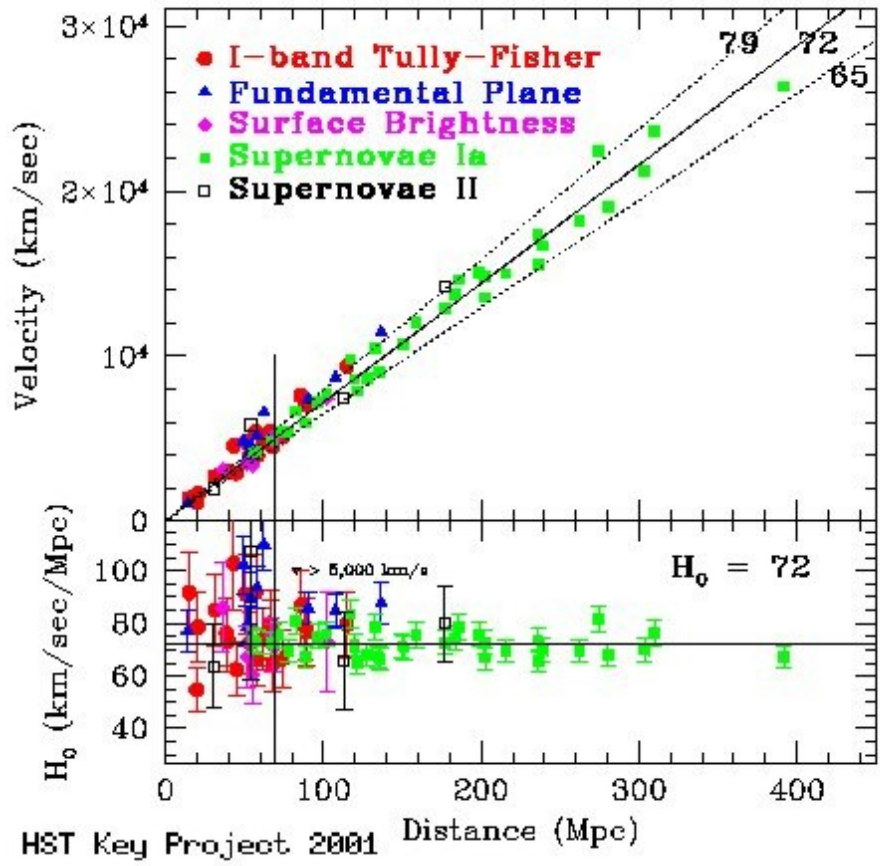

Figure 1. Summary Diagram from Freedman, et.al., 2001 for the results for the value of $H_o$, all data.

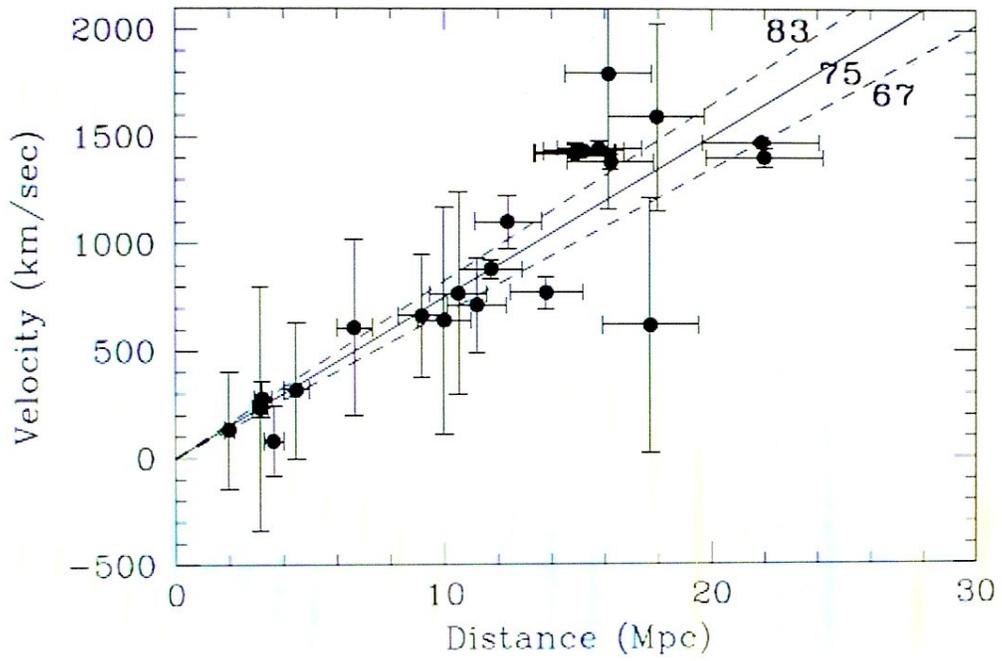

Figure 2.  Diagram from Freedman, et.al., 2001, for the results of the value of $H_o$ for the Cepheids.

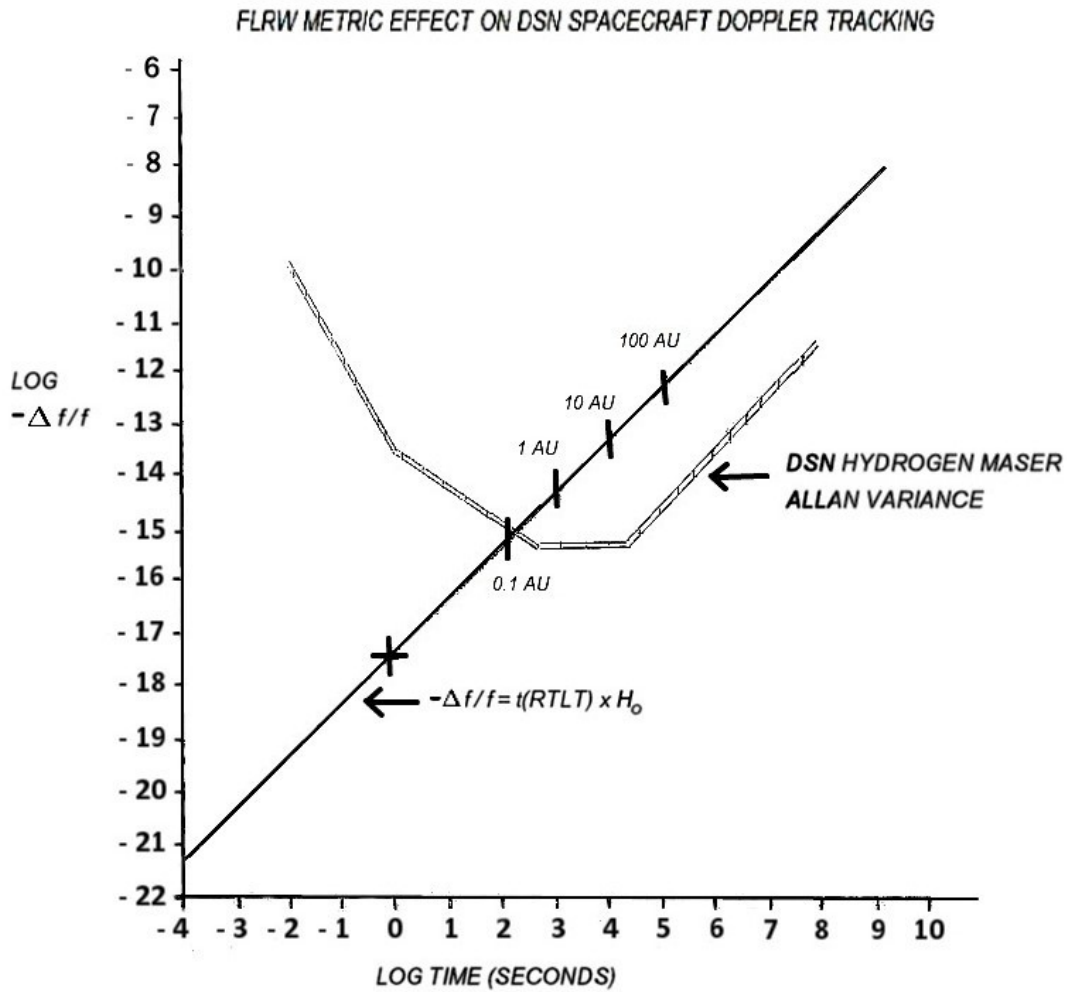

Figure 3. The effect of the FLRW metric on the Doppler tracking signal of the DSN. The values in AU are given for the full 2-Way Doppler signal equal to the return trip light time (RTLT). The ODP converts this value into a 1-Way equivalent Doppler for the analysis, however the threshold for coherence of the FLRW effect in the data is reached by the 2-Way estimate. This implies that the FLRW metric effect is incorporated in the DSN tracking data from about a distance of 0.1 AU and beyond depending upon the particular frequency standard in use at the time. The closed loop DSN Doppler system will conserve the data in the form of $H_o$ as $s^{-1}$ marked as the cross (+) on the line.

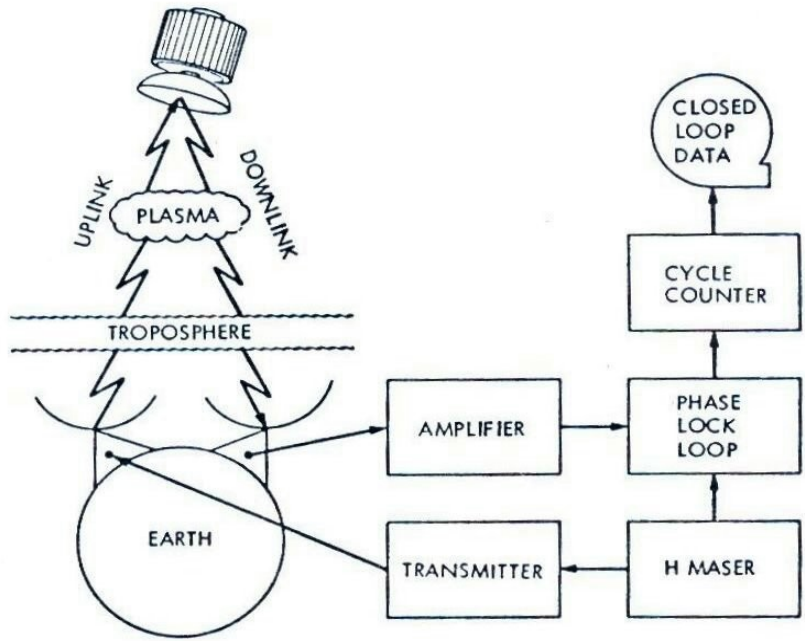

Figure 4.  Simple Schematic of the DSN Spacecraft Doppler tracking system (from ref. (10)).

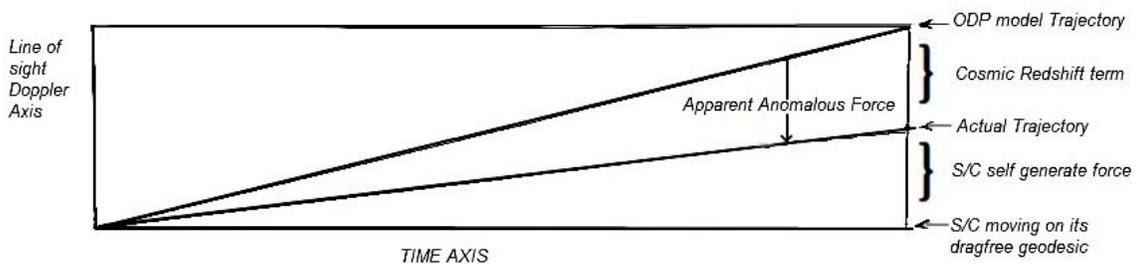

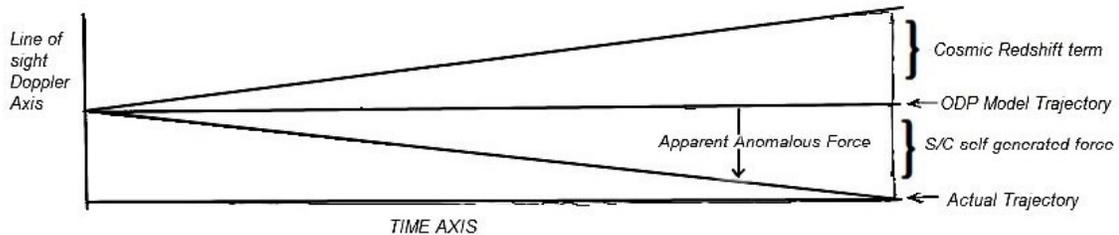

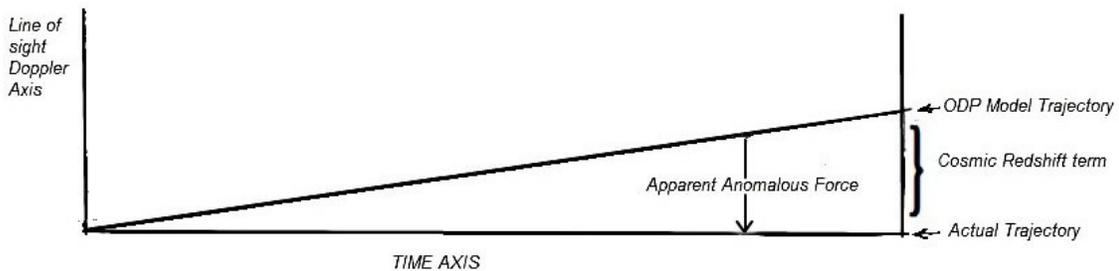

Figure 5. Three Examples of how an underdetermined ODP orbit solution represents the Cosmic Redshift term. This demonstrates that the Doppler residual from an underdetermined ODP orbit solution for the Cosmic Redshift term is independent of the spacecraft (S/C) non gravitational (self generated) force term. The Cosmic Redshift term is uniquely separated from the actual trajectory and generates Doppler residuals of the same equivalent value with reverse sign.

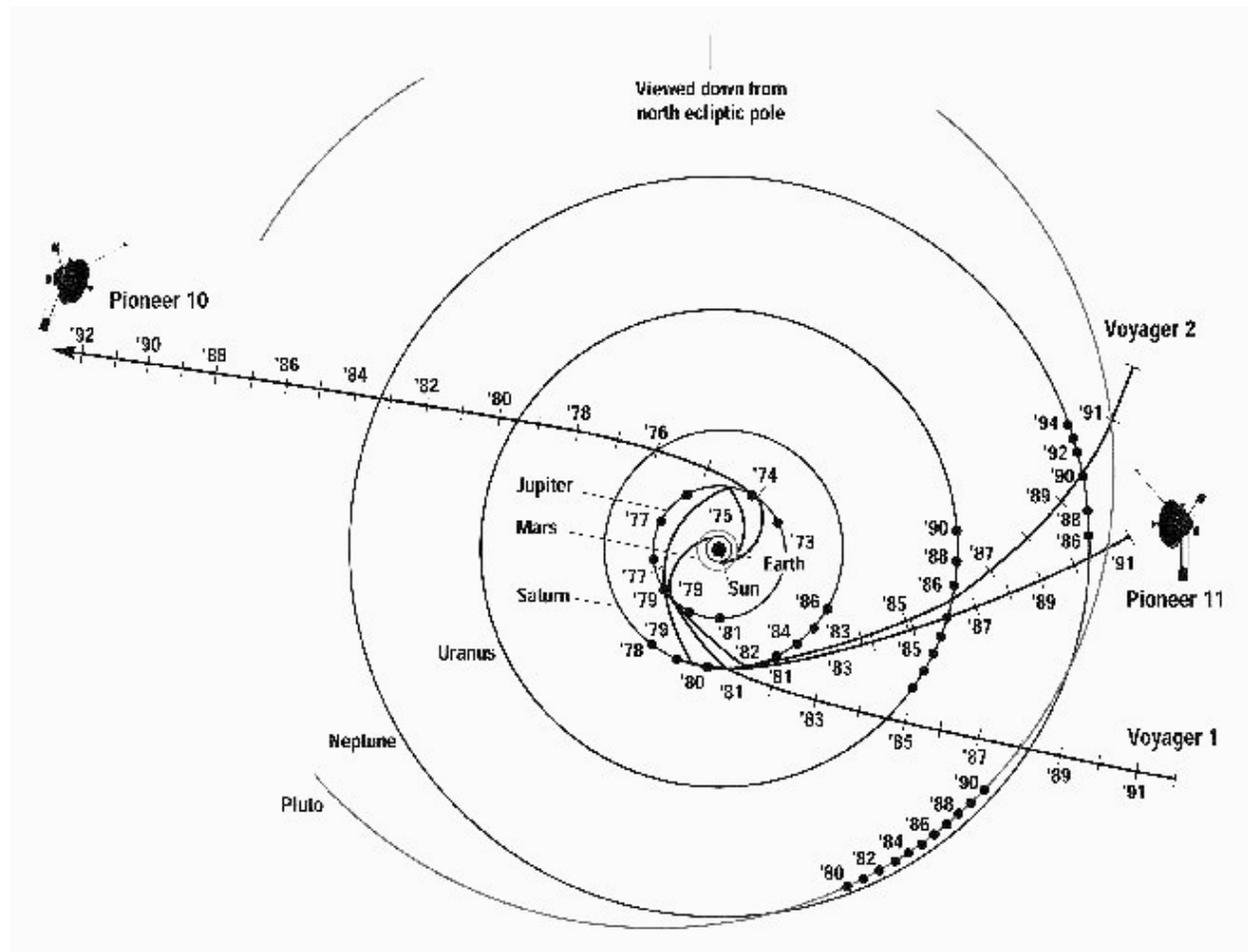

Figure 6.  Schematic of trajectories of Pioneer 10 & 11 and Voyager 1 & 2. Tracking data of this study is from the years 1987 to 1999 extending into a period beyond this figure. (Figure 3 from paper (4)).

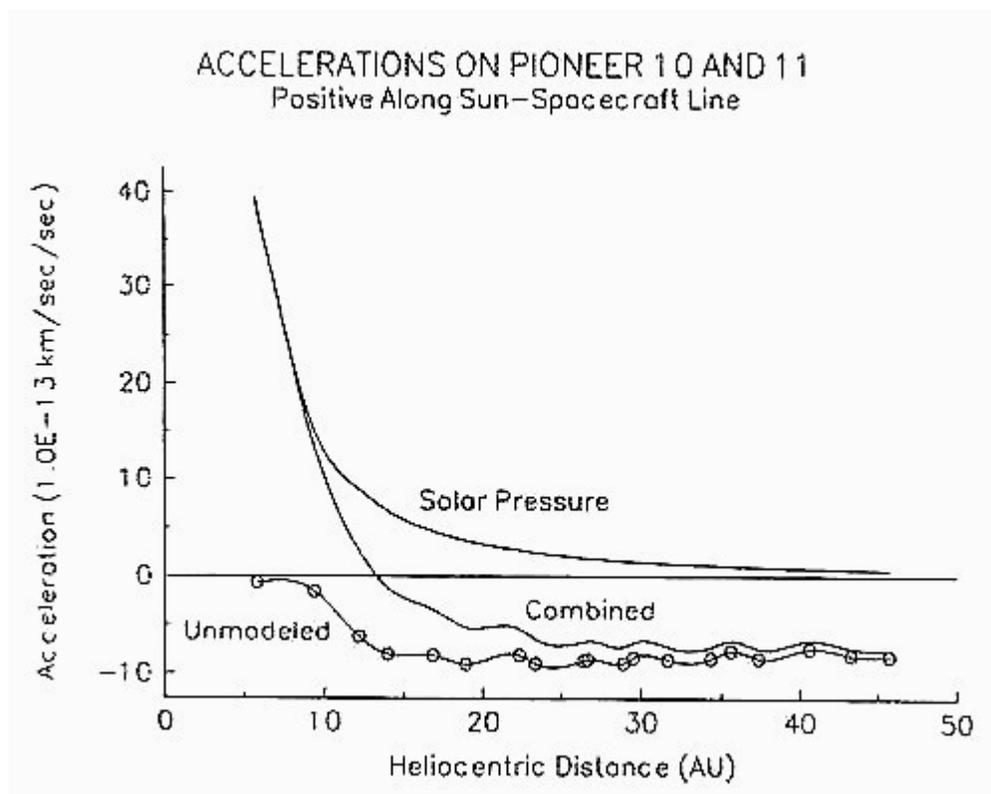

Figure 7.  Diagram shows that the "anomalous force" emerges in the Doppler residuals as the solar radiation pressure term diminishes. (Figure 6 from paper (4)).

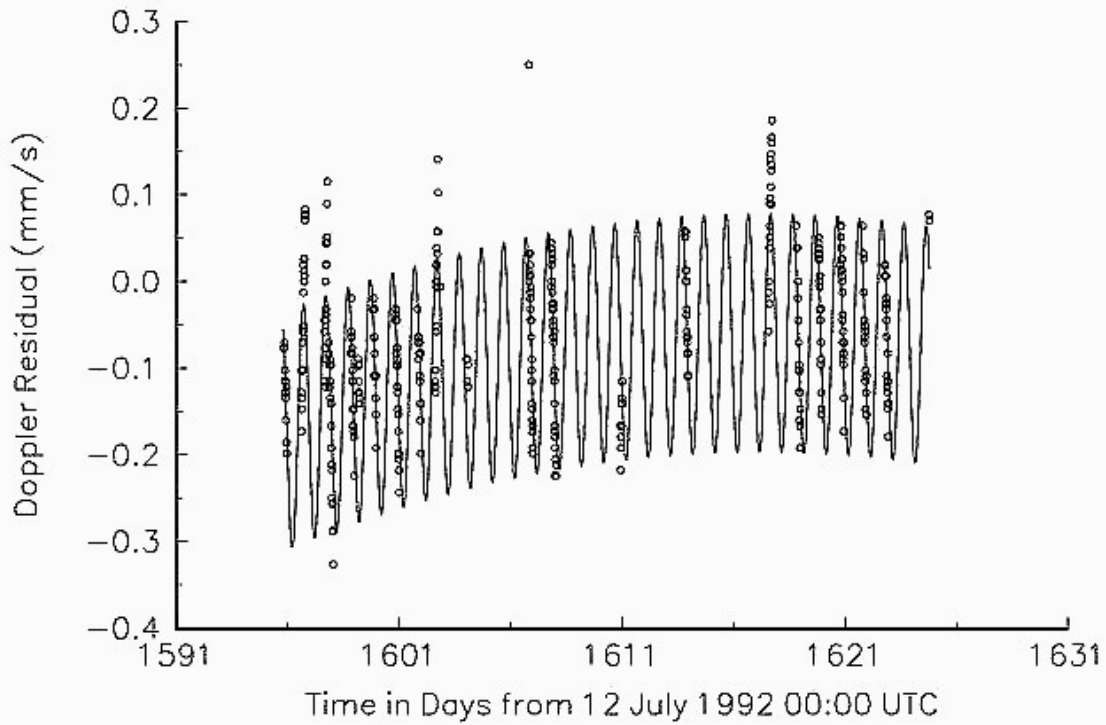

Figure 8. Detail of the Doppler residuals showing the annual and diurnal terms in the data. (Figure 18 from paper (4)). This implies the orbit of the Pioneer spacecraft is poorly determined. These terms have little effect on the determination of the linear frequency drift of the Cosmic Redshift term and are included in the variance of the 11 ½ year 1-day batch sequential solution.

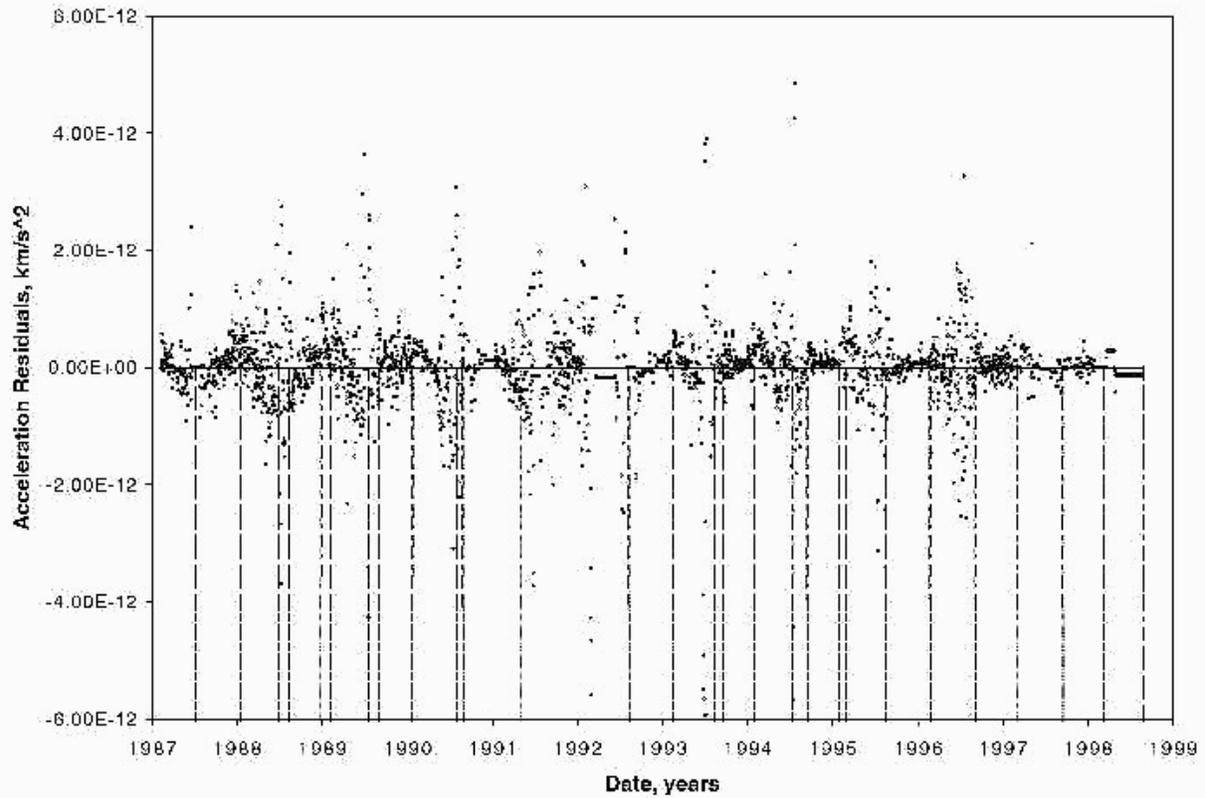

Figure 9. The Orbit Determination Program (ODP) 1-day batch sequential acceleration residuals using the entire Pioneer 10 data set. Maneuver times are indicated by the vertical dashed lines . (Figure 17 from paper (4)). The Y-axis is in units of $10^{-12}$ Km/s², or as $10^{-9}$ m/s².

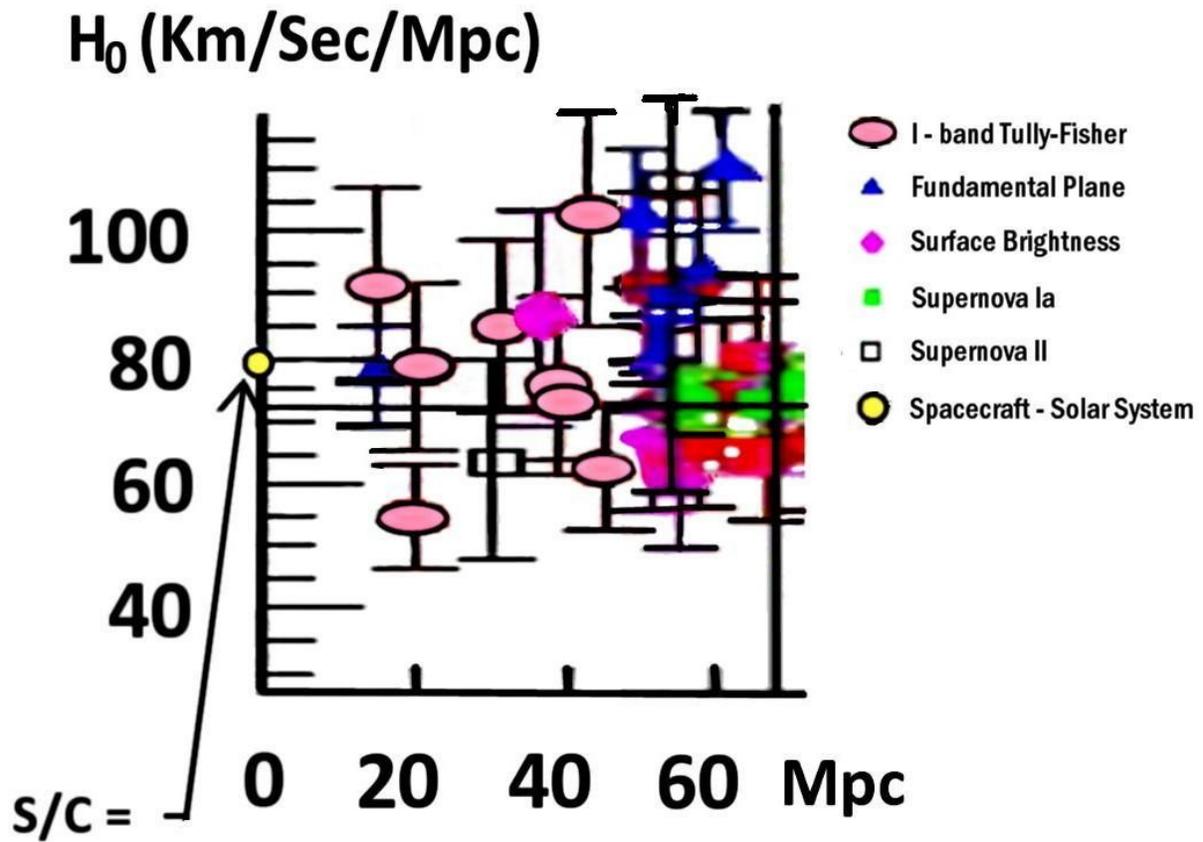

Figure 10. Detail from figure 1 of the estimates of $H_0$ for distances less than ≈ 60 Mpc from the HST Key study plotted together with the result from this study. The spacecraft result from the solar system is plotted as the yellow ball on the Y axis. The 1 sigma error bar for the spacecraft result from the solar system consistent with the other error bars of this figure is smaller in size than the yellow ball.

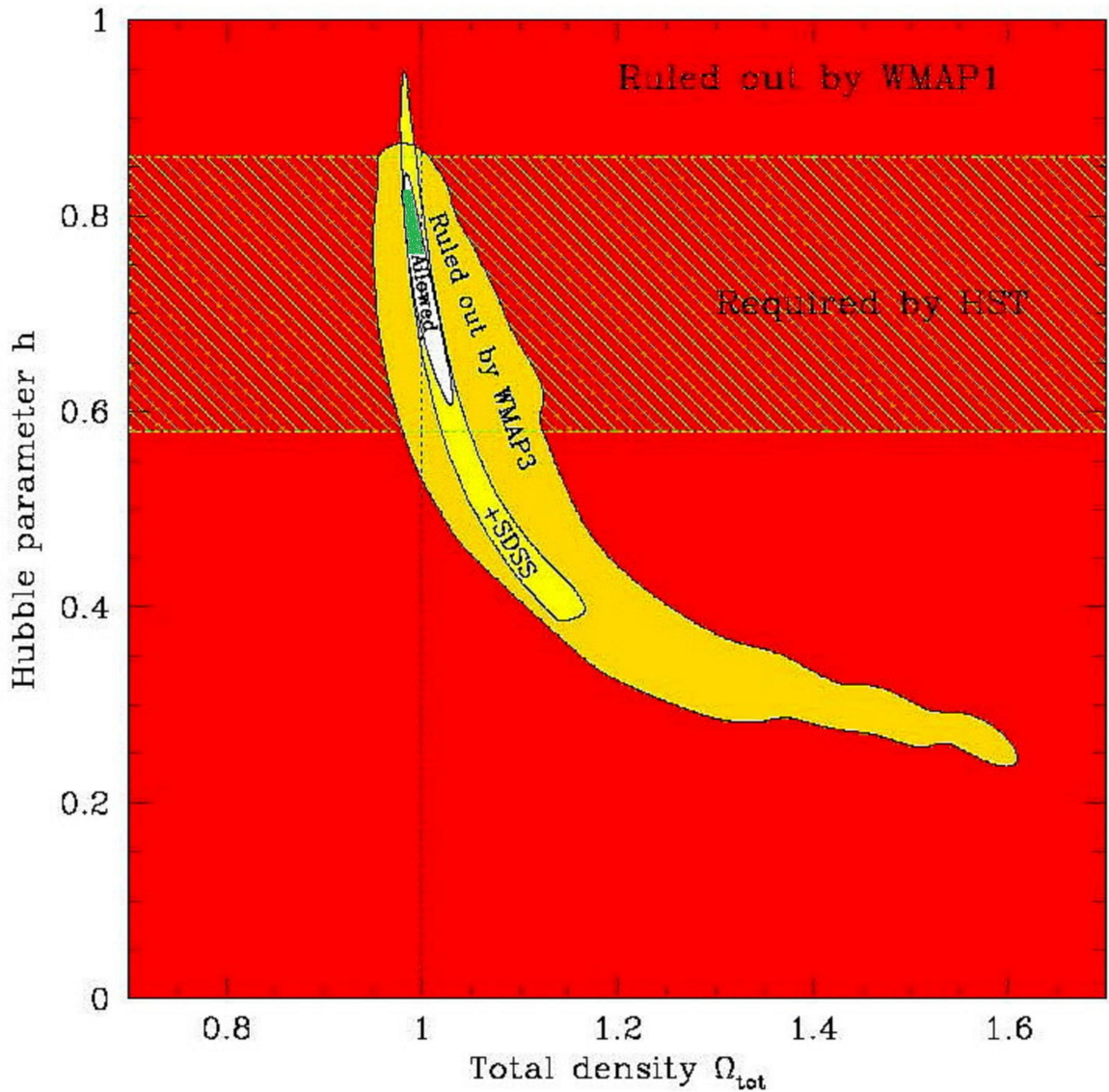

Figure 11. This figure shows the allowed region for the Hubble parameter, h, equal to $H_o/(100$ km/s/Mpc) plotted against $\Omega_{tot}$ for the λCDM model solutions of paper (12). The data is for 95% constraints in the ($\Omega_{tot}$, h) plane for 7-parameter curved models. The dotted vertical line represents $\Omega_{tot}=\Omega_{crit}=1$. The spacecraft result for the solar system using a 2 sigma error estimate consistent with the 95% model constraints of this figure is marked in green. At $\Omega_{tot}=1$ the region marked in green for $H_o$ given here is between 76.4 and 77.5. (From figure 16 of ref. (12) and this paper).